\documentclass[useAMS,usenatbib,referee]{mn2e}

\usepackage{subfigure}

\usepackage{amsmath}
\usepackage{amssymb}
\usepackage{epsfig}

\def\ggrav{\,{g_{\rm grav}}}
\def\grad{\,{g_{\rm rad}}}
\def\Gamrad{\,{\Gamma_e}}
\def\kms{\,{\rm km~s^{-1}}}
\def\mso{\,{\rm M}_\odot}
\def\rso{\,{\rm R}_\odot}
\def\msoy{\,{{\rm M}_\odot~{\rm yr}^{-1}}}
\def\vesc{\,{v_{\rm esc}}}

\def\apj{\,{ApJ}}
\def\apjs{\,{ApJS}}
\def\nat{\,{Nat}}

\title[Numerical models of continuum-driven winds]
{Numerical simulations of continuum-driven winds of super-Eddington stars}
\author[A. J. van Marle et. al.]{A. J. van Marle$^{1}$\thanks{E-mail:
marle@udel.edu (AJvM)}, S. P. Owocki$^{1}$\thanks{E-mail:
owocki@bartol.udel.edu (SPO)} and  N. J. Shaviv$^{2}$\thanks{E-mail:
shaviv@phys.huji.ac.il  (NJS)}\\
$^{1}$Bartol Research Institute, University of Delaware, 104 The Green, Newark, DE 19716, USA\\
$^{2}$Racah Institute of Physics, Hebrew University, Giv'at Ram, Jerusalem 91904 Israel}

\begin{document}

\date{Accepted ?. Received ?; in original form ?}

\pagerange{\pageref{firstpage}--\pageref{lastpage}} \pubyear{?}

\maketitle

\label{firstpage}

\begin{abstract}
We present the results of numerical simulations of continuum-driven winds of stars that exceed the Eddington limit 
and compare these against predictions from earlier analytical solutions.
Our models are based on the assumption that the stellar atmosphere consists of clumped matter, where the individual clumps 
have a much larger optical thickness than the matter between the clumps.
This `porosity' of the stellar atmosphere reduces the coupling between radiation and matter, since photons tend to 
escape through the more tenuous gas between the clumps. 
This allows a star that formally exceeds the Eddington limit to remain stable, yet produce a steady outflow from the region 
where the clumps become optically thin. 
We have made a parameter study of wind models for a variety of input conditions in order to explore the properties of 
continuum-driven winds.

The results show that the numerical simulations reproduce quite closely the analytical 
scalings.
 The mass loss rates produced in our models are much larger than can be achieved by line driving. 
 This makes continuum driving a good mechanism to explain the large mass loss and flow speeds of giant outbursts, as observed 
in {$\eta$~Carinae} and other luminous blue variable (LBV) stars.
 Continuum driving may also be important in population III stars, since line driving becomes ineffective at low metalicities.
 We also explore the effect of photon tiring and the limits it places on the wind parameters.
\end{abstract}


\section{Introduction}

Massive, hot stars continuously lose mass through radiation driving. 
The most commonly explored mechanism is line driving, wherein the scattering of photons by ions in the stellar atmosphere 
transfers momentum from the radiation field to the gas. 
 This mechanism results in a quiescent mass loss with mass loss rates ranging up to about $10^{-4}~\mso$~yr$^{-1}$ 
 \citep{so06}, which can explain the winds of most massive, hot stars.
 In fact, for most stars, the observed mass loss rates are 
 considerably less \citep{vk05}.
 
However, some stars, most notably luminous blue variable stars (LBVs) such as 
{$\eta$~Carinae}, experience  outbursts with mass loss rates several orders of 
magnitude higher than can be explained through line driving \citep{dh97, 
s02, om08}.
 In the case of $\eta$~Carinae, the 1840's outburst is inferred to
resulted in the ejection of ca. 10-20~$M_{\odot}$ over a time lasting
several years, even up to a decade 
\citep{dh97}.
While short compared to evolutionary
timescale of millions of years, this is much longer
than the typical dynamical timescale of hours, characterized by either
free-fall time  or interior sound travel time across a stellar radius.
Thus in contrast to SN ``explosions'' that are effectively driven by the
overpressure of superheated gas in the deep interior,
explaining LBV outbursts requires a more sustained mechanism that can 
drive a quasi-steady mass loss, a stellar wind, from near the stellar surface.
In the case of $\eta$~Carinae the outburst was accompanied by a strong
increase in radiative luminosity, very likely making it well above the
Eddington limit for which {\em continuum} driving by just electron scattering would
exceed the stellar gravity.
The reason for this extended increase in luminosity is not yet understood, and
likely involves interior processes beyond the scope of this paper. 
Instead, the focus here is on the way such continuum driving can result in a 
sustained mass loss that greatly exceeds what is possible through line
opacity.

A key feature of continuum driving is that, unlike line driving, it
does not become saturated from self-absorption effects in a very
dense, optically thick region. 
Indeed, since both continuum acceleration and gravity scale with the inverse square of the radius, a star that 
exceeds the Eddington limit formally becomes gravitationally unbound not only at the surface but throughout. 
Clearly, this is in contradiction to the steady surface wind mass loss observed for these stars.
N.B. Contrary to what is sometimes claimed, this would not automatically destroy the entire star. Although radiative 
acceleretion might overcome gravity locally, the total energy in the radiation field would not  suffice to drive the entire 
envelope of the star to infinity (this is known as `photon tiring'). Instead, the outward motion of the gas would quickly 
stagnate and matter would start to fall back. Nevertheless, the net result would not resemble a steady wind.

 This problem can be resolved by assuming that the stellar material is clumped rather than homogeneous, with the individual 
clumps being optically thick - and therefore self-shielding from the radiation - whereas the medium in between the clumps is 
relatively tranparent to radiation. 
 This so called `porosity effect' can lead to a reduced coupling between matter and radiation \citep{s98,s00}. 
 The photons tend to escape through the optically thin material between the clumps without interacting with the matter inside 
the clumps. 
 This implies that a star that formally exceeds the Eddington limit can remain gravitationally bound and would only exceed 
the effective Eddington limit at the radius where the individual clumps themselves become optically thin.

 The structure of such a star should therefore look as follows \citep{s01}.  
 Deep inside the super-Eddington star, convection is necessarily excited \citep{jso73} such that the radiation field remains 
sub-Eddington through most of the stellar interior. 
 At low enough densities where maximally efficient convection cannot sufficiently reduce the radiative flux, the near 
Eddington luminosity necessarily excites at least one of several possible instabilities \citep[e.g.,]{a92,s01a} which give 
rise to a reduced opacity. This `porous'  layer has a reduced effective opacity and an increased effective Eddington 
luminosity. Thus, the layer remains gravitationally bound to the star.
 At lower densities still, the dense clumps become optically thin and the effective opacity approaches its microscopic value. 
From this radius outwards, the matter is gravitationally unbound, and is part of a continuum-driven wind. 

 A detailed analytical study of this paradigm was carried out by {Owocki, Gayley \& Shaviv (2004)}, hereafter \citet{ogs04}. 
 This predicted that continuum-driven winds can produce high mass loss rates ($\geq10^{-3}~\msoy$) at intermediate wind 
velocities ($10^2-10^3~\kms$).
 Here we test these analytical predictions with numerical simulations of winds from super-Eddington stars.

 In addition to LBVs, which are the specific objects we study here, other types of astronomical objects can exceed the 
Eddington limit and therefore experience similar continuum-driven winds. 
 These include for example classical novae \citep{s01} or high accretion rate accretion disks around black holes \citep{b06}.
 In fact, classical nova eruptions clearly exhibit steady continuum-driven winds, indicating that the mass loss rate is 
somehow being regulated and likely to be described by the porosity model and the same continuum-driven wind analyzed here.

 The layout of this paper is as follows. In  \S\ref{sec-subEdd} we show the effect of continuum scattering on a 
sub-Eddington, line-driven wind.
 In  \S\ref{sec-analytic} we summarize the analytic results obtained by \citet{ogs04}. 
 In \S\ref{sec-numeric} we describe the numerical methods that we have used for our simulations.
 \S\ref{sec-result} shows the results of our simulations and the comparison with the analytical predictions.
 In \S\ref{sec-phtir} we discuss the effect of photon tiring and show how it influences the results of our calculations.
 Finally, in \S\ref{sec-disc} we end with a summary and a discussion.
\begin{figure*}
 \centering
\resizebox{\hsize}{!}{\includegraphics[width=\textwidth,angle=-90]{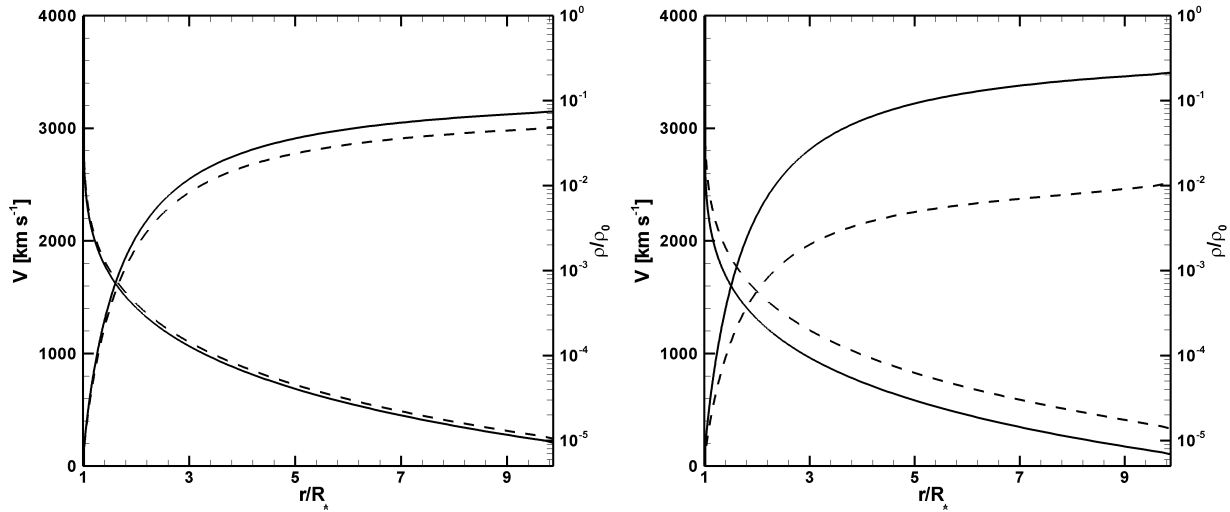}}      
 \caption{
 Left: The effect of continuum driving on a sub-Eddington, line-driven wind for $\Gamrad=0.1$. The solid lines depict the 
wind parameters for the CAK force only, the dashed lines show the joint effect of both CAK line driving and continuum 
scattering. Clearly, the influence of the continuum term is negligeable.
 Right: The same variables , but for $\Gamrad=0.5$. Here the difference caused by continuum driving is significantly more 
pronounced.}
 \label{fig:line1}
 \end{figure*}

  \section{Sub-Eddington limit}
  \label{sec-subEdd}
  As long as a star does not exceed the Eddington limit, continuum interaction alone cannot drive a stellar wind. 
  However, it does influence the parameters of a (primarily) line-driven wind. 
  Figure~\ref{fig:line1} demonstrates the influence of continuum driving on the stellar wind parameters. 
  For these simulations we use the CAK line driving formalism described in \citet{do02}, including both the finite disk 
correction factor and the modified gravity scaling of the line force (eq.~[19] of \citet{ogs04}). For the same stellar 
parameters ($50~\mso$, $25~\rso$, 50\,000~K) and CAK parameters (${\bar{Q}=2.0\times10^3}$,
${\alpha=2/3}$) but two different 
Eddington parameters ($\Gamrad=L_\star/L_{\rm Edd}=0.1$ and 0.5,) 
we simulate the stellar wind with and without the continuum 
driving term. 

Note that unlike the purely continuum-driven winds in the next section, these simulations do not contain any porosity effect. 
  Namely, the continuum driving term in the equation of motion is simply:
\begin{equation}
g_{\rm cont}~=~ \kappa_e \frac{F_{\rm rad}}{c},
\end{equation}
 with the continuum opacity $\kappa_e$ set to $0.3$~g/cm$^2$.
 {\bf These parameters give us an Eddington luminosity of ${L_{\rm Edd}=4\pi GcM/\kappa_e=8.33\times10^{39}}$~erg/s.}
 Figure~\ref{fig:line1} shows both the density (relative to the density at the stellar surface) and wind velocity as a 
function of radius for these two simulations. 
 Clearly, for a low Eddington factor, the effect of continuum interaction on the wind parameters is negligeable. 
 However, once the star approaches the Eddington limit, continuum driving increases the mass loss rate and decreases the 
terminal velocity.
For $\Gamrad=0.1$ the mass loss rate increases from $1.80\times10^{-6}\msoy$ to $1.92\times10^{-6}\msoy$ 
because of continuum driving. For $\Gamrad=0.5$, these numbers are $2.18\times10^{-6}\msoy$ and $3.22\times10^{-6}\msoy$ 
respectively.
\section{Analytical approximation}
\label{sec-analytic}
\subsection{Wind velocity}
 For a radiatively driven outflow from a stellar surface, irrespective of the specific driving mechanism involved, the 
equation of motion can be written as
 \begin{equation}
 \biggl[1-\frac{a^2}{v^2}\biggr] v\frac{dv}{dr}~=~-\frac{G M_\star}{r^2} + \grad + \frac{2a^2}{r}- \frac{{\rm d}a^2}{{\rm 
d}r},
 \label{eq:mot1}
 \end{equation}
 with $a$ the isothermal sound speed, $\grad$ the radiative acceleration and ${G M_\star}/{r^2}$ the inward gravitational 
acceleration. 
 The last two terms on the right hand side are the result of the gas pressure gradient. 
 In most cases they be neglected as their contribution to the velocity is usually negligeable next to the gravitational and 
radiative acceleration terms \citep{ogs04}. 
 This assumption becomes incorrect only in those situations where the terminal velocity of the wind is close to the sound 
speed.
 If we neglect these gas pressure terms, we can rewrite equation (\ref{eq:mot1}) in a dimensionless form, 
 \begin{equation}
 \biggl[1-\frac{w_s}{w}\biggr]w'~=~-1+\Gamrad,
 \label{eq:mot2}
 \end{equation}
 with $w=(v/\vesc)^2$, $w_s=(a/\vesc)^2$ and $\Gamrad=\grad r^2/GM_\star$. 
 Note that $\vesc$ is defined as the escape velocity at the surface of the star $R_\star$ rather than as a local escape 
velocity. 
 The gravitationally scaled inertial acceleration,
 \begin{equation}
 w'~=~\frac{r^2 v {\rm d}v/{\rm d}r}{GM_\star},
 \end{equation}
 can be written in terms of the inverse radius coordinate $x\equiv1-R_\star/r$, so that $w'={\rm d}w/{\rm d}x$. 
 This inverse radius coordinate makes for a more practical coordinate system than the radius itself, since it scales
 with the gravitational potential.

Since the escape velocity of a massive star is typically at least an order of magnitude larger than the local 
isothermal sound speed, we can in many cases neglect the bracketed term.
For the porous atmosphere, \citet{ogs04} introduced a dimensionless parameter $k_b=\kappa_{\rm eff}/\kappa$, 
which gives the 
effective opacity of the local clumped medium vs.\ the opacity of the medium if it were completely homogeneous.
 Like \citet{ogs04}, we assume that the distribution $(f)$ of optical depths of the individual clumps can be described by a 
truncated power law,
 \begin{equation}
 \tau \frac{{\rm d}f}{{\rm d}\tau}~=~\frac{1}{\Gamma(\alpha_p)}\biggl(\frac{\tau}{\tau_0}\biggr)^{\alpha_p}   
e^{-\tau/\tau_0},
 \end{equation}
 with $\alpha_p \geq 0$ being the power-index and $\tau_0$ the optical depth of the {\it thickest} clump. 
 Here $\Gamma(\alpha_p)$ is the gamma function, rather than the Eddington parameter.
 This gives us:
 \begin{equation}
 k_b~=~\frac{(1+\tau_0)^{1-\alpha_p} - 1}{(1-\alpha_p)\tau_0}.
 \end{equation}
 For the full derivation see \citet{ogs04}. 
 
 As long as the clump characteristics are fixed, the optical depth $\tau_0$ scales with the density as, 
 \begin{equation}
 \tau_0 = \rho/\rho_0, 
 \end{equation}
 where $\rho_0$ is the critical density at which the thickest blob has unit optical depth. 
 This is calculated by
 \begin{equation}
 \rho_0~=~\frac{1}{\kappa h_0},
 \end{equation}
 with $\kappa$ the continuum opacity and $h_0$ the porosity length of the thickest clump; the porosity length $h$ is defined 
as $L^3/l^2$, with $L$ being the typical separation between clumps and $l$ the size of the clump.

We thus rewrite eq.~(\ref{eq:mot2}) by substituting,
 \begin{equation}
 \Gamrad \longrightarrow \Gamrad k_b(\tau_0),
 \end{equation}
for the Eddington parameter $\Gamrad=L_\star/L_{\rm Edd}$.
The scaled velocity gradient thus takes the form,
 \begin{equation}
 w'(x)~=~ \frac{\Gamrad k_b [\tau_0(x)] -1}{1-w_{s}/w}
 \, .
 \label{eq:w}
 \end{equation}
This is a first-order differential equation that can be integrated using standard numerical
techniques, starting from the sonic-point initial condition $w(x=0)=w_{s}$, 
where the vanishing of both the numerator and denominator on the right-hand side 
requires application of L'Hopital's rule to evaluate the initial gradient, 
\begin{equation}
w'_{s} \equiv w'(x=0) = \frac{\alpha_{p}}{4} \, 
\left [ 1 + \sqrt{1+32 w_{s}/\alpha_{p}} \right ]
\, .
\label{eq:wps}
\end{equation}

By introducing the porosity term into the equation of motion, we have
made the radiative acceleration dependent on the density and therefore
on the radius.
 Under these circumstances, the radiative acceleration is only able to drive matter away from the star 
in the outermost layers of the star, 
producing a steady wind that can last as long as the star remains
about the Eddington limit.
In effect, porosity {\em regulates} the mass loss to a level that can be
sustained in a nearly steady way throughout an LBV outburst.
Here again the word ``outburst'' has to be interpreted in context. 
LBV outbursts can last years, which, though short compared to stellar evolutionary time, 
is quite long compared to a SN explosion. 
Most importantly, this is much longer
than a typically dynamical flow time (ca. a free fall time, $R/\vesc$, or a
wind expansion time, $R/v_\infty$).  
As such, the mass loss can be well modeled in
terms of steady-state wind solutions that assume the luminosity, etc. are
constant over such dynamical timescales.

\subsection{Mass loss rate}
 The mass loss rate induced by continuum driving is directly related to the luminosity of the star and the power index 
$\alpha_p$. 
 The generic mass loss rate that follows from the porosity is
 \begin{equation}
 \dot{M}~=~\biggl(\frac{\rho_\star}{\rho_0}\biggr)\biggl(\frac{H}{h_0}\biggr)
\frac{L_{\rm Edd}}{ac},
\label{eq:mdotbasic}
 \end{equation}
with $H$ the gravitational scale height of the gas at the sonic point.
The sonic point density,  $\rho_\star$,  can be found through condition
 \begin{equation}
 \Gamrad k_b~=~1.
\label{eq:gamkbeq1}
 \end{equation}
For the simple case $\alpha_{p} = 1/2$, this has the explicit solution
\begin{equation}
    \biggl(\frac{\rho_\star}{\rho_0}\biggr) 
    = 4 \Gamrad \, (\Gamrad - 1)  ~~~ ; ~~~ \alpha_{p} = \frac{1}{2} 
\end{equation}
which leads to an explicit expression for the mass loss rate
(cf. eqn. (77) in \cite{ogs04}), 
\begin{equation}
 \dot{M} = 4(\Gamrad-1)\frac{L_\star}{\eta_0 a c} ~~~ ; ~~~ \alpha_p=\frac{1}{2} 
 \, ,
\label{eq:mdot}
\end{equation}
with ${\eta_0 \equiv h_0/H}$.
For more general cases, \citet{ogs04} also give explicit
approximations for the expected mass loss rate, but in quoting values 
for this ``analytic porosity model'' below, we choose the more
accurate approach of solving  eqn.~(\ref{eq:gamkbeq1}) implicitly, and
using the resulting $\rho_{\star}$ to compute the associated mass loss rate
from eq.~(\ref{eq:mdotbasic}).

\section{Numerical method}
\label{sec-numeric}
 For our numerical simulations we use the ZEUS hydrodynamics code \citep{sn92, c96}. 
 We do our computations on a 1D spherical grid with an inflow boundary in the center and outflow at the outer boundary. 
Continuum driving is modeled through an acceleration term $\Gamrad k_b \ggrav$ 
added to the radial component of the equation of motion (see also, \citet{mos08a}).

 We start our simulation by initializing a 1D radial grid with matter moving away from the origin with a power law velocity 
distribution. 
 At the inner boundary, we specify an inflow density, higher than the critical density $\rho_s$ to ensure that the sonic 
point (where the wind velocity equals the isothermal sound speed) is inside the grid. 
 This inflow density remains constant during the simulation. 
 The inflow velocity is time-dependent, and recalculated at each timestep such that the velocity gradient over the boundary 
is equal to the velocity gradient immediately above the boundary radius.
 Such ``floating'' boundary conditions allow approach to a stable,
steady, flow  solution, and have been successfully used for 
line-driven-wind simulations in both one \citep{ocr88} and two dimensions \citep{ocb94}.

 The radial grid is not evenly spaced, but rather chosen in such a fashion that the individual grid size at the sonic point 
is always smaller than the local scale height.
 
 For our simulations we assume that the gas is isothermal with a temperature of 50\,000\,K, which implies an isothermal sound 
speed of approximately $20~\kms$. 
 The opacity of the gas is set to 0.4 gr/cm$^2$ (Thompson scattering opacity for pure hydrogen) and $\eta_0$ to 1.
 The radius and mass of the central star are set to 50~$\rso$ and 50~$\mso$ respectively.
 {\bf A star such as this has an Eddington luminosity of $6.24\times10^{39}$~erg/s.}

 The grid extends from the stellar surface to a distance of ten times the stellar radius.

 \begin{figure}
 \centering
\resizebox{\hsize}{!}{\includegraphics[width=\textwidth]{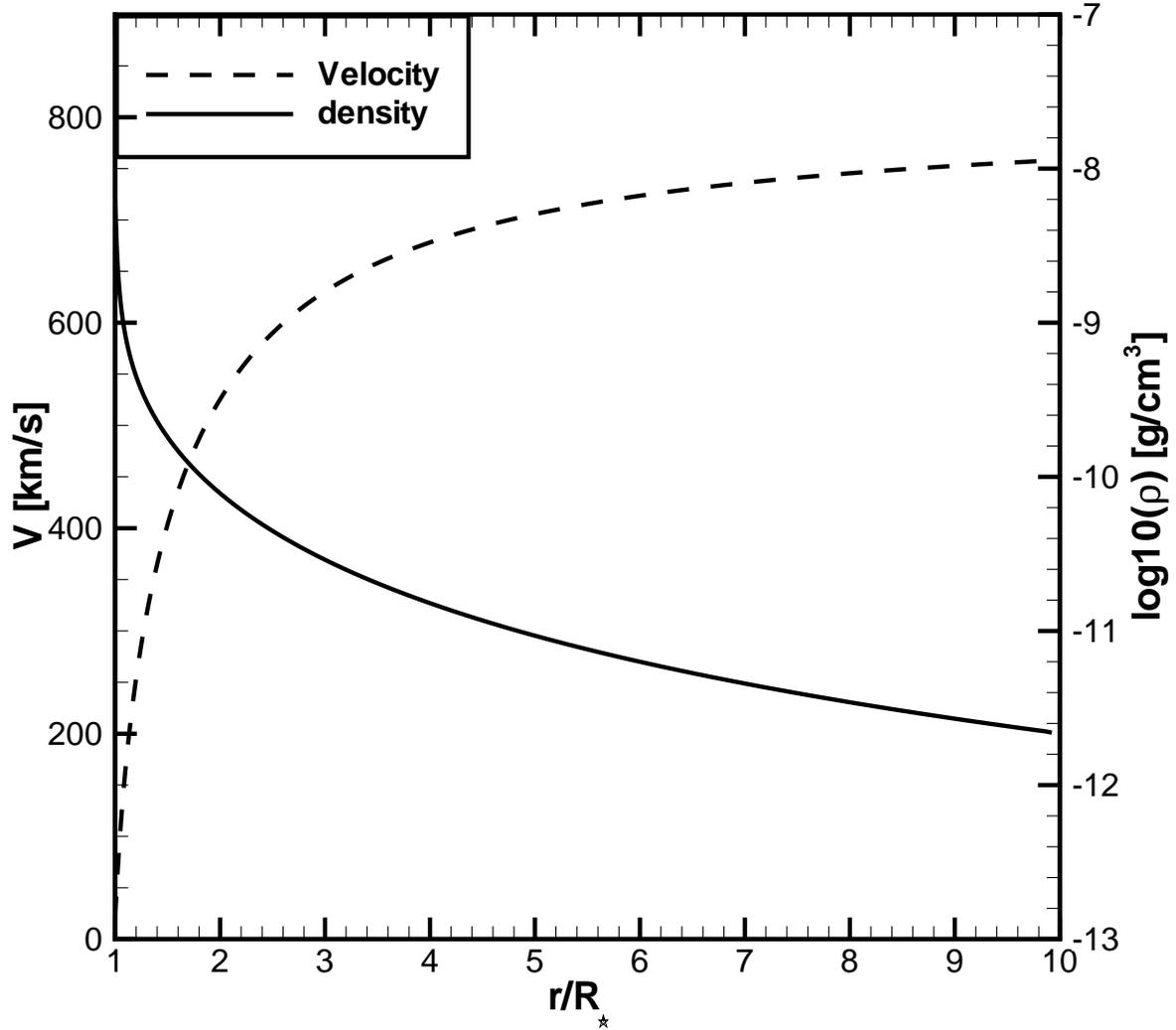}} 
 \caption{Wind velocity and density as a function of radius for a continuum-driven wind. On the horizontal axis, the local 
radius divided by the sonic point radius. Independent parameters: $\Gamrad=3$, $\alpha_p=0.5$, $w_\star=0.001$}
 \label{fig:windprofile}
 \end{figure}     
 
 \begin{figure}
 \centering
\resizebox{\hsize}{!}{\includegraphics[width=\textwidth]{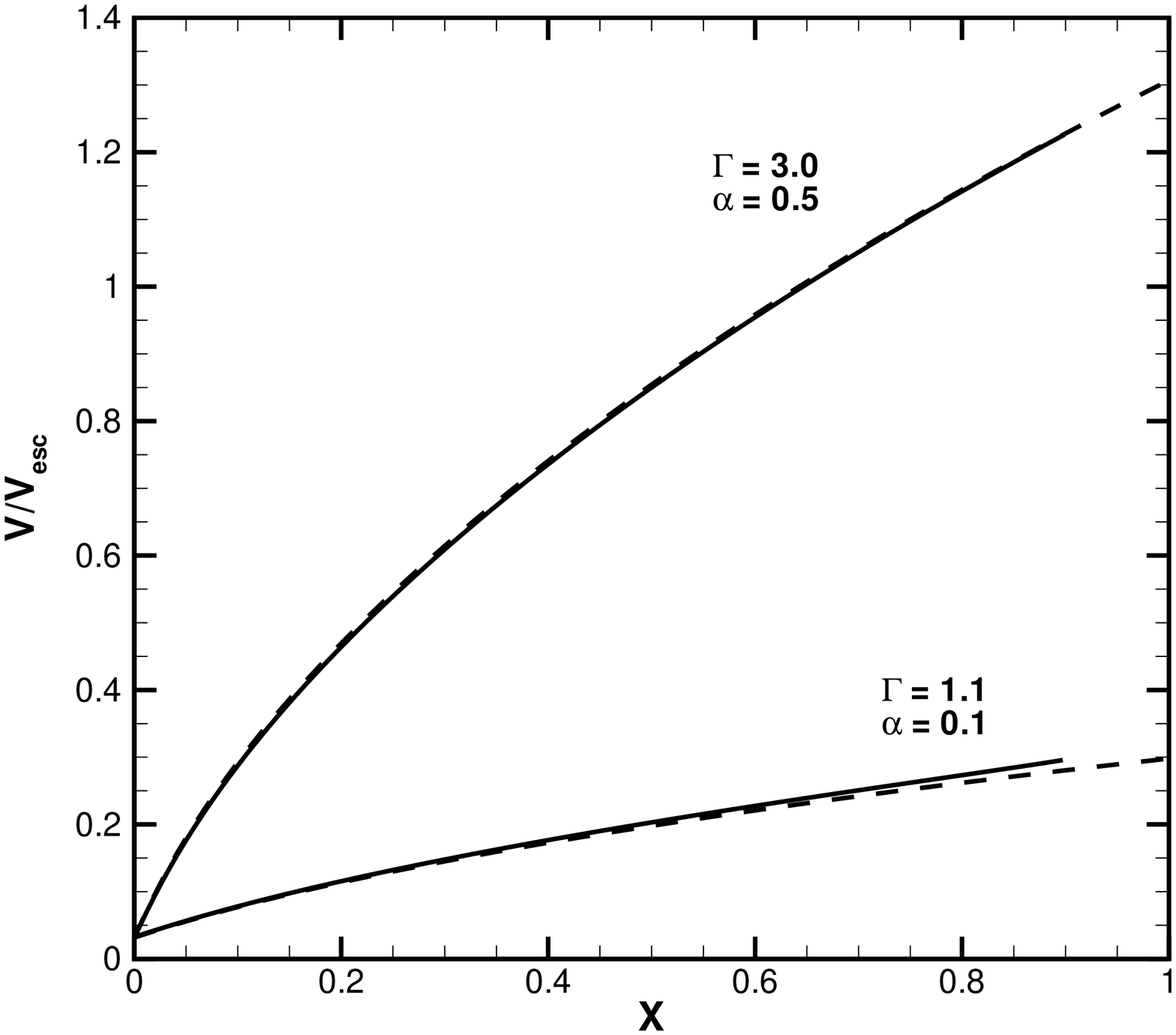}}      
 \caption{ 
Wind velocity in units of the sonic point escape radius plotted versus
the dimensionless radial coordinate $x=1-R_{sonic}/r$,
comparing results for the analytical solution (dashed curves) and numerical
simulation (solid curves) for a continuum-driven wind.
The upper curves show results for the  same parameters as
figure~\ref{fig:windprofile}, with $\Gamrad=3$ and $\alpha_p=0.5$,
while the lower curves are for the marginal super-Eddington case
$\Gamrad=1.1$, with $\alpha_{p}=0.1$.
The agreement between numerical and analytical solutions is quite
good, with only a slightly stronger differences for the $\Gamrad=1.1$ 
case.
}
 \label{fig:vel12}
 \end{figure} 


\section{Results}
\label{sec-result}
 A grid of 30 models was calculated, encompassing a parameter space of $0.1\leq\alpha_p \leq 0.8$ and $1.1\leq\Gamrad \leq 
6.0$. 
 Stellar mass, radius surface temperature, opacity and clumping parameter $\eta_0$ were kept constant.
 A typical result for these calculations is shown in figure~\ref{fig:windprofile}, which shows the density and wind velocity 
as a function of distance from the star.

\clearpage
   \begin{table*}
   \centering
	  \begin{tabular}{p{1cm} p{3cm} p{2cm} p{2cm} c r}
	 \hline
	 \hline
	 \noalign{\smallskip}
	     $\Gamrad$  & $\alpha_p$ & $\dot{M}_{an}$  & $\dot{M}_{num}$ & $\dot{M}_{an}/\dot{M}_{num}$ &v/$\vesc$  \\
		      &            & $\mso yr^{-1}$      & $\mso yr^{-1}$       &           \\
	 \noalign{\smallskip}
	 \hline \hline
	  6.0  &  0.8    &    4.86$\times 10^{-2}$       &  4.87$\times 10^{-2}$ & 0.997 & 1.93  \\
	  6.0  &  0.5    &    1.98$\times 10^{-1}$       &  1.99$\times 10^{-1}$ & 0.997 & 1.80  \\
	  6.0  &  0.3    &    2.09                       &  2.10                 & 0.994 & 1.50  \\
	  6.0  &  0.2    &    3.92$\times 10^{1}$        &  3.99$\times 10^{1}$  & 0.982 & 1.13  \\
	  6.0  &  0.1    &    2.87$\times 10^{5}$        &  3.00$\times 10^{5}$  & 0.956 & 0.655 \\
	 \hline 
	  4.0  &  0.8    &    2.43$\times 10^{-2}$       &  2.44$\times 10^{-2}$ & 0.995 & 1.53  \\
	  4.0  &  0.5    &    7.94$\times 10^{-2}$       &  7.97$\times 10^{-2}$ & 0.996 & 1.47  \\
	  4.0  &  0.3    &    5.23$\times 10^{-1}$       &  5.28$\times 10^{-1}$ & 0.991 & 1.31  \\
	  4.0  &  0.2    &    5.13                       &  5.23                 & 0.981 & 1.08  \\
	  4.0  &  0.1    &    4.97$\times 10^{3}$        &  5.21$\times 10^{3}$  & 0.954 & 0.655 \\
	 \hline
	  3.0  &  0.8    &    1.42$\times 10^{-2}$       &  1.42$\times 10^{-2}$ & 0.997 & 1.26 \\
	  3.0  &  0.5    &    3.97$\times 10^{-2}$       &  4.00$\times 10^{-2}$ & 0.992 & 1.23 \\
	  3.0  &  0.3    &    1.91$\times 10^{-1}$       &  1.93$\times 10^{-1}$ & 0.990 & 1.14 \\
	  3.0  &  0.2    &    1.20                       &  1.22                 & 0.985 & 1.00 \\
	  3.0  &  0.1    &    2.80$\times 10^{2}$        &  2.92$\times 10^{2}$  & 0.959 & 0.654 \\
	 \hline
	  2.0  &  0.8    &    5.83$\times 10^{-3}$       &  5.92$\times 10^{-3}$ & 0.984 & 0.907 \\
	  2.0  &  0.5    &    1.32$\times 10^{-2}$       &  1.34$\times 10^{-2}$ & 0.987 & 0.894 \\
	  2.0  &  0.3    &    4.20$\times 10^{-2}$       &  4.27$\times 10^{-2}$ & 0.983 & 0.862 \\
	  2.0  &  0.2    &    1.47$\times 10^{-1}$       &  1.50$\times 10^{-1}$ & 0.980 & 0.812 \\
	  2.0  &  0.1    &    4.83                       &  5.06                 & 0.955 & 0.633 \\
	 \hline
	  1.5  &  0.8    &    2.53$\times 10^{-3}$       &  2.57$\times 10^{-3}$ & 0.984 & 0.643 \\
	  1.5  &  0.5    &    4.96$\times 10^{-3}$       &  5.09$\times 10^{-3}$ & 0.974 & 0.643 \\
	  1.5  &  0.3    &    1.19$\times 10^{-2}$       &  1.22$\times 10^{-2}$ & 0.978 & 0.632 \\
	  1.5  &  0.2    &    2.87$\times 10^{-2}$       &  2.95$\times 10^{-2}$ & 0.973 & 0.616 \\
	  1.5  &  0.1    &    2.61$\times 10^{-1}$       &  2.72$\times 10^{-1}$ & 0.958 & 0.551 \\
	 \hline
	  1.1  &  0.8    &    4.33$\times 10^{-4}$       &  4.88$\times 10^{-4}$ & 0.888 & 0.301\\
	  1.1  &  0.5    &    7.27$\times 10^{-4}$       &  8.10$\times 10^{-4}$ & 0.898 & 0.300\\
	  1.1  &  0.3    &    1.32$\times 10^{-3}$       &  1.48$\times 10^{-3}$ & 0.892 & 0.299\\
	  1.1  &  0.2    &    2.20$\times 10^{-3}$       &  2.47$\times 10^{-3}$ & 0.891 & 0.298\\
	  1.1  &  0.1    &    6.08$\times 10^{-3}$       &  6.81$\times 10^{-3}$ & 0.892 & 0.293\\
      \hline \hline
   \end{tabular}
   \caption{ For parameters $\Gamrad$ and $\alpha_{p}$ given in the
   first two columns, comparison of the mass loss rates predicted by
   the analytic porosity model (column 3) vs. those obtained in full
   numerical simulations (column 4), with the fifth column giving the 
   ratio, analytical/numerical. The agreement is remarkably good, but with
   modest, ca. 10\% differences for lower $\alpha_{p}$ or $\Gamrad$
   near unity. The last column gives the ratio between the terminal velocity and 
   the escape speed from the sonic radius for the numerical models.}
   \label{tab:mdot}
   \end{table*}

Table~\ref{tab:mdot} compares mass loss rates for the numerical and analytical solutions for
 a sample of model parameters $\Gamrad$ and $\alpha_{p}$.
As noted above, the analytical mass loss rate is computed from 
eqn.~(\ref{eq:mdotbasic}) using the sonic point density derived from 
implicit solution of eqn.~(\ref{eq:gamkbeq1}).
The comparison shows that the analytical and numerical results coincide very
well, with maximum differences of about 10\% for $\Gamrad$ near unity 
and small $\alpha_{p}$. 

The wind velocity can be approximated semi-analytically by integrating equation~(\ref{eq:w}). 
The results again match the numerical simulation quite well, as plotted in
figure~\ref{fig:vel12}.

The match is partcularly good for the standard parameter set used in figure \ref{fig:windprofile}, with
$\Gamrad=3$ and $\alpha_{p}=0.5$.
For the marginal super-Eddington
model $\Gamrad=1.1$ with small power index $\alpha_{p} = 0.1$, 
which represents an almost pathogical case, 
the differences are greater, but still quite modest.

\begin{figure*}
\centering
\mbox{\subfigure{\includegraphics[width=0.45\textwidth,angle=-90]{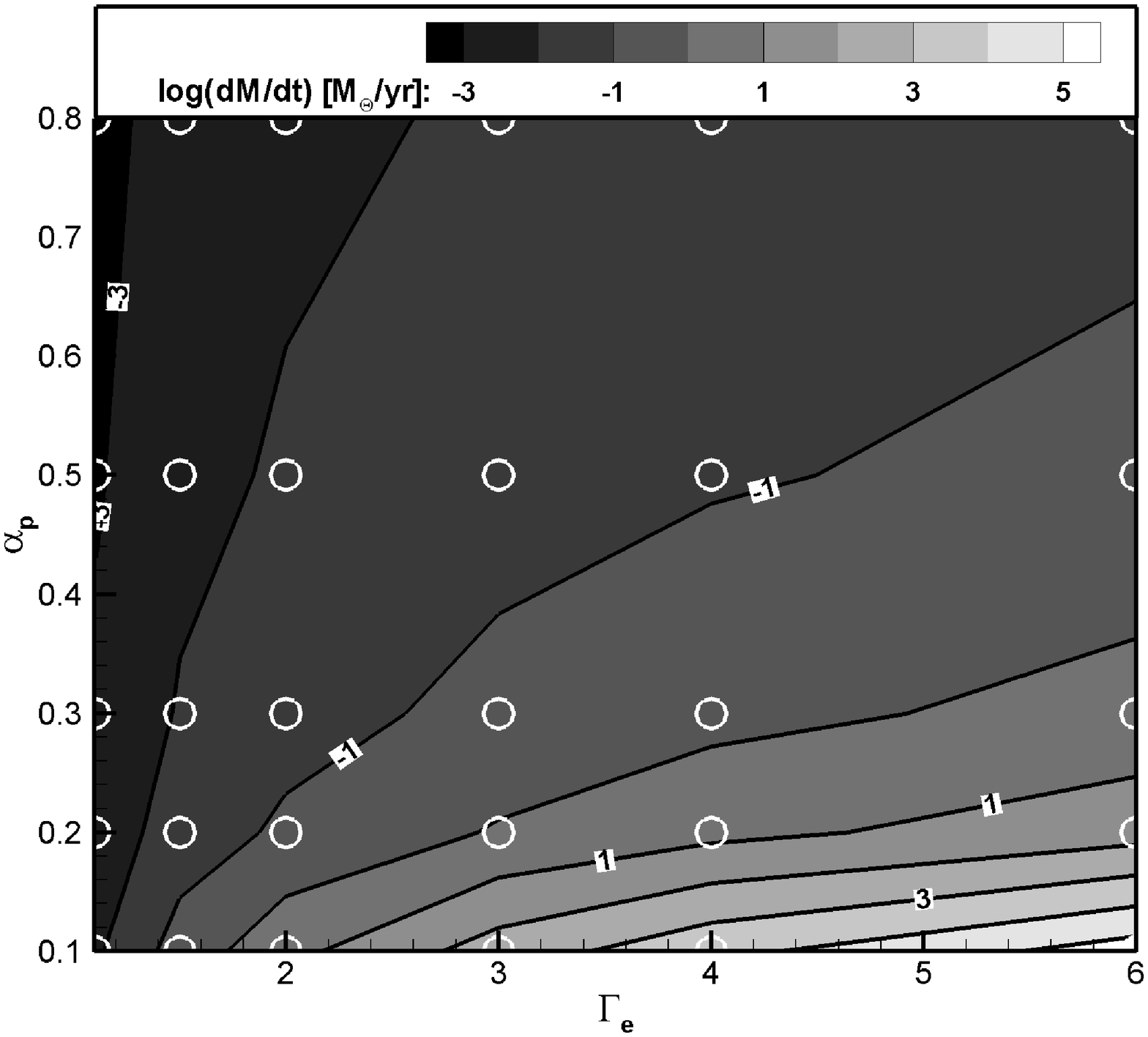}}
      \subfigure{\includegraphics[width=0.45\textwidth,angle=-90]{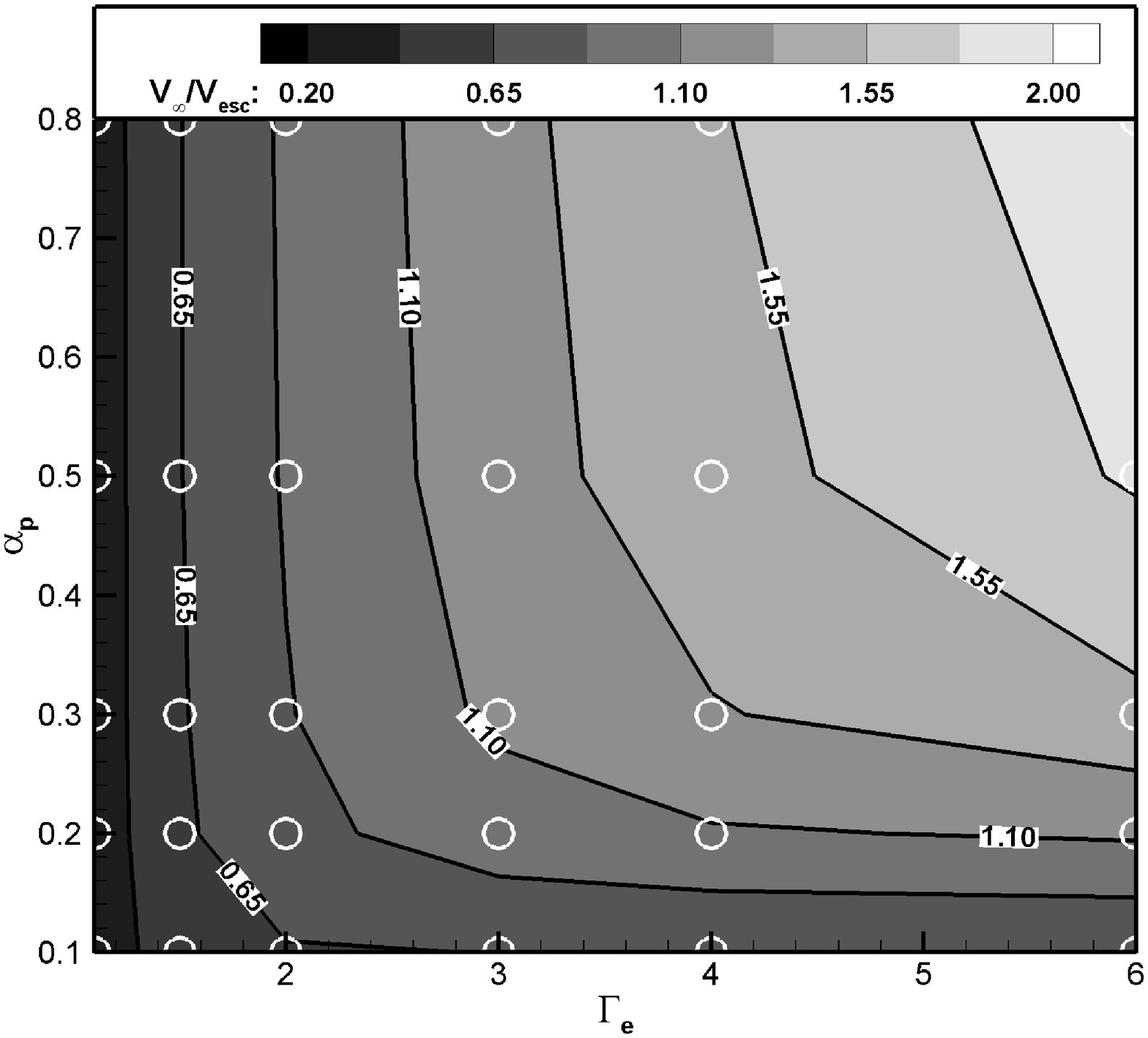}}}
\caption{
Left:
Total mass loss rate in solar masses per year as a function of $\Gamrad$ and $\alpha_p$. As predicted in the analytical 
approximations (see \S\ref{sec-analytic}) the mass loss rate increases with $\Gamrad$ but decreases with $\alpha_p$. 
Right:
Terminal wind velocity divided by escape velocity as a function of $\Gamrad$ and $\alpha_p$. The wind velocity increases with 
both $\Gamrad$ and $\alpha_p$.
The white circles indicate the actual datapoints, taken from table~\ref{tab:mdot}.
}
\label{fig:dM}
\end{figure*}

The complete results of the parameter study are shown in fig.~\ref{fig:dM}, which shows the mass loss rate and terminal 
wind velocity for all simulations. 
As expected, the mass loss rate increases with $\Gamrad$ and decreases with $\alpha_p$ as predicted in 
\S~\ref{sec-analytic}. 
The terminal wind velocity increases both with $\Gamrad$ and with $\alpha_p$. 
This is hardly surprising.
A high $\Gamrad$ means a strong radiative acceleration, which will lead to a higher velocity.
A high $\alpha_p$ means a weaker coupling between matter and radiation.  
This decreases the mass loss rate, which for a given total available energy means an increase in velocity.  

Mass loss rates in general are very high, though one should remember that photon-tiring was not included in these 
simulations. 
Although photon tiring does not change the mass loss rate at the stellar surface, it can prevent that part of the mass 
escapes the stellar gravity.
Wind velocities are usually on the order of the escape velocity.

\section{Photon tiring}

One of the most interesting aspects of continuum-driven winds is their high mass loss rates, which gives rise to the 
phenomenon of ``photon tired winds" suggested by \citet{og97}. That is, winds in which the energy required to drive the gas 
out of the gravitational well is comparable or even larger than the energy available in the radiation field. In 
\S\ref{sec-tired-analytic} we briefly cover the analytical description of these winds, and then continue with a numerical 
study of the case in which there is sufficient energy in the radiation field. The opposite limit, which gives rise to 
non-stationary solutions will be dealt in the subsequent publication. 

\label{sec-phtir}
\subsection{Analytical approximation}
\label{sec-tired-analytic}
 If the stellar wind is driven by radiation, conservation of energy limits the maximum mechanical luminosity of the star to 
the radiative luminosity. 
 This effect is known as photon tiring and was not included in the simulations described in the previous section. 

 The maximum amount of mass that can be driven of by radiation is the mass loss rate for which the mechanical luminosity of 
the wind matches the radiative luminosity, 
 \begin{equation}
 \frac{1}{2}\dot{M}\vesc^2~\leq~L_{\rm tir}~=~L_{\star},
 \end{equation}
 with $L_\star$ the radiative luminosity of the star.
 This sets the upper limit for the mass loss rate to,
 \begin{equation}
 \dot{M}_{\rm tir}~=~\frac{2 L_\star}{\vesc^2}.
 \end{equation}
 Note that this is the same for both line driving and continuum driving, and is unrelated to the composition of the stellar 
atmosphere. 
 It is a fundamental property of the star. 
 From this limit we calculate the photon tiring number~$m$,
 \begin{equation}
 m~=~\frac{\dot{M}}{\dot{M}_{\rm tir}}.
 \label{eq:mtir}
 \end{equation}
 As long as ${m \ll 1}$, photon tiring has no significant influence on the wind properties, but when ${m \lesssim 1}$ the 
situation changes. 
 The radiation field of the star becomes depleted as much of its energy is used to drive the matter, leaving less energy 
available to drive the outer layers of the wind. 
 The mass loss rate at the surface remains the same since it is set by the local radiative acceleration, but the wind 
velocity will decrease. 

 This will also influence observations, since radiative energy that is used to drive the wind will no longer be part of the 
visible radiation output of the star. 
 For stars with line-driven winds, this is hardly worth considering. 
 Even a Wolf-Rayet star with a powerful wind loses only a few percent of its luminosity to the wind. For continuum-driven 
winds, the effect can become quite significant.

 Should $m$ actually exceed unity, the radiation field loses all its energy. 
 As a result, the matter that leaves the stellar surface will not be able to reach the escape velocity and start to fall back 
onto the star.
 This will make it impossible to obtain a steady state solution. 

 A analytic analysis of the photon tiring effect was done by \cite{ogs04}, predicting that while the mass loss rate would 
remain the same, the velocity of the wind would drop. 
 The new equation for the wind velocity becomes:
\begin{equation}
 w'(x)~=~\frac{\kappa_{\rm eff}[\tau_0(x)]}{\kappa} \Gamrad[1-m(w+x)]  - 1.
\label{eq:wtiring}
\end{equation}  
 Note that $\Gamrad$ is no longer a constant but decreases with the radius and the velocity. 
 The photon tiring number can be found by combining equations (\ref{eq:mdot}) and (\ref{eq:mtir}). 
 For the special case where $\alpha_p=1/2$ this implies that,
\begin{equation}
 m~=~0.13\frac{\Gamrad -1}{\eta_0 a_{20}}\frac{M_\star}{\mso}\frac{\rso}{R_\star},
\end{equation}    
 with $a_{20}$ the sound speed in units of $20~{\rm km/s}$.
 The implications of this equation are quite clear. 
 The relative effect of photon tiring increases with $\Gamrad$ and decreases with the porosity length. 
 This is to be expected, as a larger $\Gamrad$ means an increase in mass loss rate, and therefore an increase in the amount 
of energy necessary to lift the material. 
 A larger $\eta_0$ means a decrease in coupling between radiation and matter, such that the effect of photon tiring 
diminishes. 
 The same is true for an increase in $\alpha_p$, so we can expect the effect of photon tiring to diminish with higher 
$\alpha_p$.

\subsection{Numerical method}
 We calculate the effect of photon tiring by calculating the work integral along the radial gridline and subtracting the 
result from the total luminosity of the star.
 This provides us with the luminosity available at each radial grid point, which can then be used to accelerate the wind 
during the next time step.
 
 The work integral is given by:
\begin{equation}
\begin{aligned}
  W~&=~\int_{R\star}^r \dot{m}(r) \grad {\rm d}r, \\
    &=~4\pi\int_{R\star}^r \rho(r) r^2 v(r)  \grad {\rm d}r.
\end{aligned}
\end{equation}
 Therefore, the effective luminosity at a given radius equals:
\begin{equation}
\begin{aligned}
 L~&=~L_\star - W, \\
   &=~L_\star - 4\pi\int_{R\star}^r \rho(r) r^2 v(r)  \grad {\rm d}r
\end{aligned}
\end{equation}
 \citep{mos08a}.
 Using this value for the luminosity, rather than the stellar luminosity used in the simulations described in 
\S\ref{sec-result}, a new set of simulations was made for the same range of $\Gamrad$ values and $\alpha_p=0.5$. 
 All other parameters were kept the same as in \S\ref{sec-result}.

 \begin{figure}
 \centering
\resizebox{\hsize}{!}{\includegraphics[width=\columnwidth]{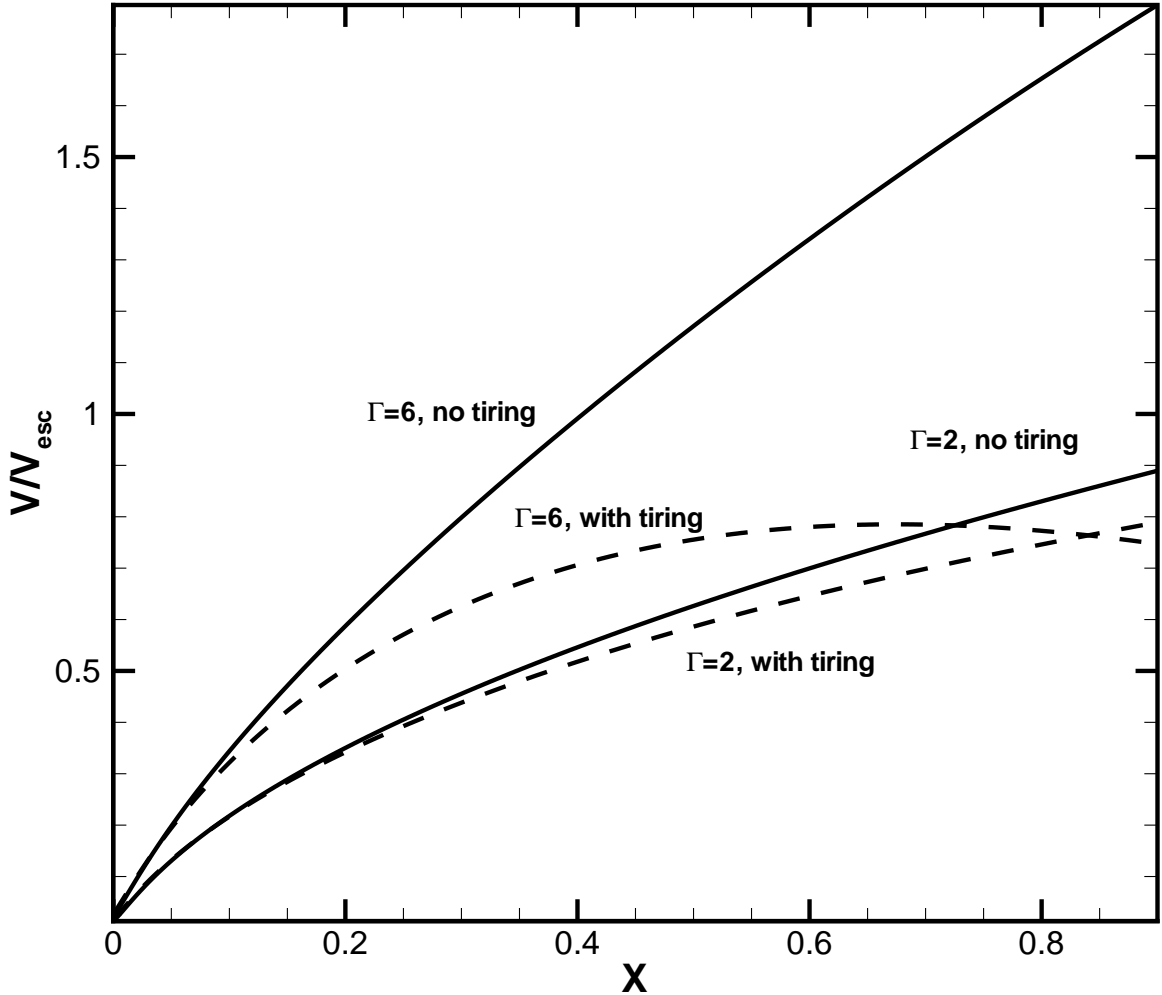}}      
 \caption{The effect of photon tiring on the wind velocity (I). Clearly, the effect is much larger for larger $\Gamrad$.}
 \label{fig:vtir}
 \end{figure}    

 \begin{figure}
 \centering
\resizebox{\hsize}{!}{\includegraphics[width=\columnwidth]{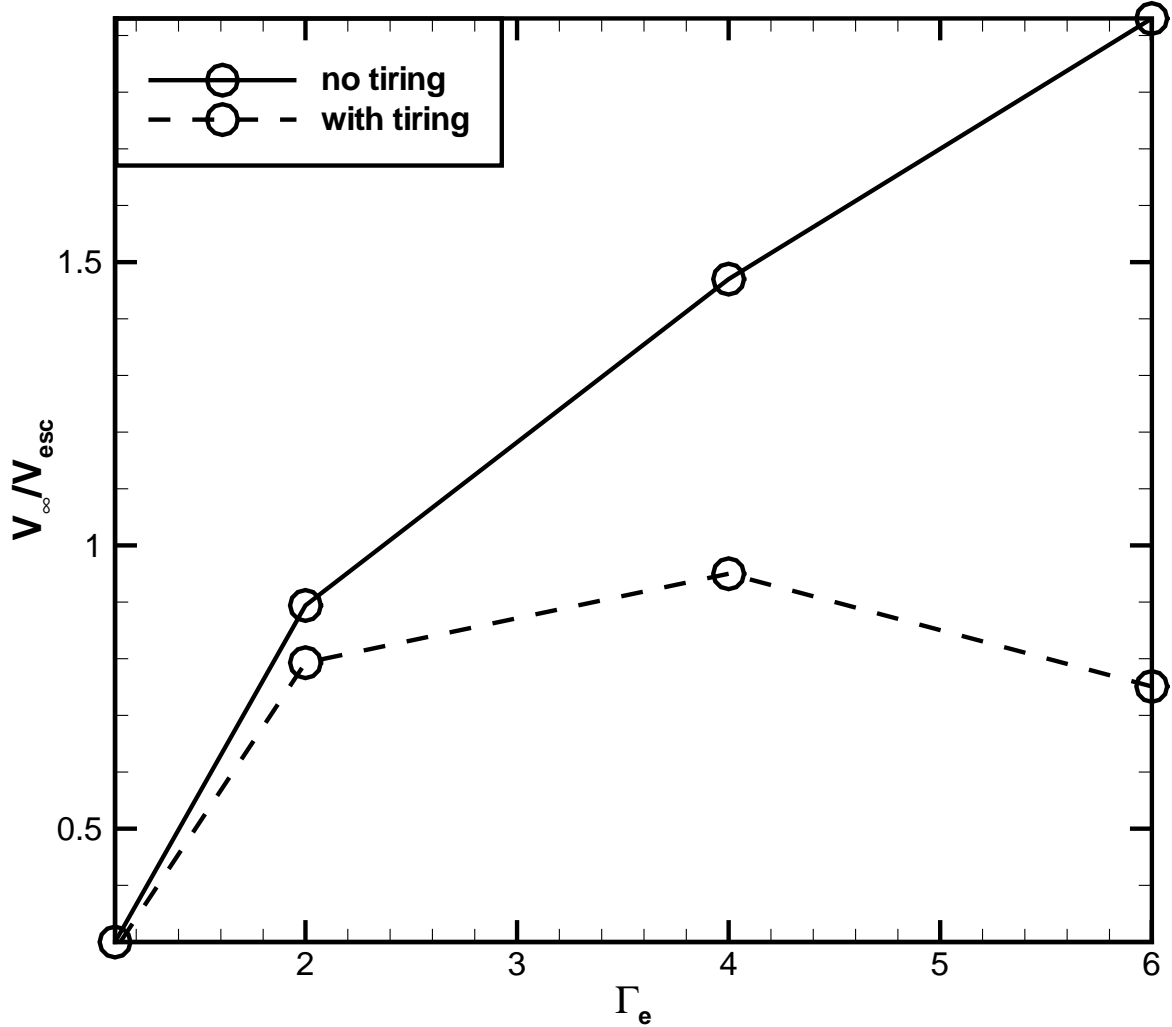}}      
 \caption{The effect of photon tiring on the wind velocity (II). This graph depicts the terminal velocity as a function of 
$\Gamrad$ for simulations with and without photon tiring included. The effect clearly increases with 
$\Gamrad$.}
 \label{fig:vtir2}
 \end{figure}    

\subsection{Numerical results}
 The effect of photon tiring is demonstrated in figure~\ref{fig:vtir}. 
 For $\Gamrad=2$, the effect is small, whereas for $\Gamrad=6$ the effect is very large. 
 Also, for the larger value of $\Gamrad$, the velocity decreases beyond a certain radius. 
 This corresponds to the radius where the luminosity falls below the Eddington luminosity.
 Note that the mass loss rate is independent of the tiring parameter and will therefore not change if photon tiring is 
included. 
 This implies that the density of the wind increases, since the velocity decreases due to the photon tiring effect.
 It therefore also implies that the optical depth of the wind is higher as well.

 In figure~\ref{fig:vtir2}, one can see a comparison between the values of the terminal velocity obtained with and without 
photon tiring.
 The influence of photon tiring clearly increases with $\Gamrad$.
 For large $\Gamrad$, the terminal velocity itself actually decreases.
 It is clear from these results that for larger values of $\Gamrad$ the photon tiring effect must be included in the 
numerical simulations, or the results will quickly become unphysical.

\section{Discussion}
\label{sec-disc}
 A series of numerical simulations was carried out in order to test the analytical approximations for the porosity length 
formalism of continuum driving, published by \cite{ogs04}.
 The numerical results coincide well with the analytical results and demonstrate that this mechanism allows for powerful 
radiation-driven winds. 
 This effect can explain the mass loss rates and wind velocities observed in Luminous Blue Variables such as  
{$\eta$~Carinae}. 

 The simulations also confirm that the photon tiring effect plays an important role in continuum driving as it places an 
upper limit on the mass loss rate. 
 The effects of actually crossing the photon tiring limit have not yet been explored. 
 This situation is much more complicated, since the simulations will no longer be able to reach a steady state solution.
  Ideally, such simulations should be done in two, or even three dimensions, to investigate the effect of 
 interactions between different layers of the stellar wind as they move back and forth.
 Note that at this point we cannot predict how the star itself would react to such a situation. 
 All our simulations have been done under the assumption that stellar parameters do not change significantly over time.  
 It is possible that conditions in the outer layers of the star would change to reduce the driving force.

  Porosity reduced continuum driving can also be important for the winds of other super-Eddington objects such as Novae 
\citep{s01}, accretion disks \citep{b06} and transients like {M85OT2006-1} \citep{ketal07}.

 In a companion paper, \citep{mos08b}, we explore the situation where the star exceeds the photon tiring limit. 
 We also intend eventually to generalize the simulations to multiple dimensions and explore the influence of stellar rotation 
on continuum-driven winds.

\section*{Acknowledgments}

A.J.v.M. acknowledges support from NSF grant AST-0507581 and N.J.S. the  
support of ISF grant 1325/06.  We thank A.~ud-Doula and R.~Townsend for helpful discussions and comments.

\bsp

\label{lastpage}

\end{document}